\documentclass[twocolumn,showpacs,preprintnumbers,amsmath,amssymb,prl]{revtex4}

\usepackage{graphicx}
\usepackage{dcolumn}
\usepackage{bm}

\newcommand{\pizero}{\pi^0}
\newcommand{\klong}{K_L^0}

\newcommand{\kzpinu}{K_L^0 \rightarrow \pi^0 \nu \bar{\nu}}

\newcommand{\kzpipi}{K_L^0 \rightarrow \pi^0 \pi^0}
\newcommand{\kzpipipi}{K_L^0 \rightarrow \pi^0 \pi^0 \pi^0}
\newcommand{\kzgg}{K_L^0 \rightarrow \gamma \gamma}

\newcommand{\pt}{P_T}
\newcommand{\zvtx}{Z_{\textrm{vtx}}}

\newcommand{\stat}{_{\textrm{(stat.)}}}

\newcommand{\syst}{_{\textrm{(syst.)}}}

\newcommand{\sigeff}{\varepsilon_{A}}

\newcommand{\nklong}{(1.67 \pm 0.04\stat ) \times 10^9}

\newcommand{\sesfinal}{(9.11 \pm 0.20\stat \pm 0.64\syst ) \times 10^{-8}}
\newcommand{\brfinal}{2.1 \times 10^{-7}}

\begin{document}

\title{New limit on the $\kzpinu$ decay rate\\}

\newcommand*{\PUSAN}{%
$^1$Department of Physics, Pusan National University, Busan, 609-735 Republic of Korea}
\newcommand*{\SAGA}{%
$^2$Department of Physics, Saga University, Saga, 840-8502 Japan}
\newcommand*{\DUBNA}{%
$^3$Laboratory of Nuclear Problems, Joint Institute for Nuclear Research, 
Dubna, Moscow Region, 141980 Russia}
\newcommand*{\SOKENDAI}{%
$^{4}$Department of Particle and Nuclear Research, 
The Graduate University for Advanced Science (SOKENDAI), Tsukuba, Ibaraki, 305-0801 Japan}
\newcommand*{\TAIWAN}{%
$^{5}$Department of Physics, National Taiwan University, Taipei, Taiwan 10617 Republic of China}
\newcommand*{\KEK}{%
$^{6}$Institute of Particle and Nuclear Studies, 
High Energy Accelerator Research Organization (KEK), Tsukuba, Ibaraki, 305-0801 Japan}
\newcommand*{\OSAKA}{%
$^{7}$Department of Physics, Osaka University, Toyonaka, Osaka, 560-0043 Japan }
\newcommand*{\YAMAGATA}{%
$^{8}$Department of Physics, Yamagata University, Yamagata, 990-8560 Japan}
\newcommand*{\CHICAGO}{%
$^{9}$Enrico Fermi Institute, University of Chicago, Chicago, Illinois 60637, USA }
\newcommand*{\NDA}{%
$^{10}$Department of Applied Physics, National Defense Academy, Yokosuka, Kanagawa, 239-8686 Japan}
\newcommand*{\RCNP}{%
$^{11}$Research Center of Nuclear Physics, Osaka University, Ibaragi, Osaka, 567-0047 Japan}
\newcommand*{\KYOTO}{%
$^{12}$Department of Physics, Kyoto University, Kyoto, 606-8502 Japan\\ \rm (E391a collaboration) }

\author{
J.K.~Ahn$^1$, 
Y.~Akune$^2$, 
V.~Baranov$^3$, 
M.~Doroshenko$^{4, a}$, 
Y.~Fujioka$^2$, 
Y.B.~Hsiung$^5$, 
T.~Inagaki$^6$, 
S.~Ishibashi$^2$, 
N.~Ishihara$^6$, 
H.~Ishii$^7$, 
T.~Iwata$^8$, 
S.~Kobayashi$^2$, 
S.~Komatsu$^7$, 
T.K.~Komatsubara$^6$, 
A.S.~Kurilin$^3$, 
E.~Kuzmin$^3$, 
A.~Lednev$^{9, b}$, 
H.S.~Lee$^1$, 
S.Y.~Lee$^1$, 
G.Y.~Lim$^6$, 
T.~Matsumura$^{10}$, 
A.~Moisseenko$^3$, 
H.~Morii$^{12}$, 
T.~Morimoto$^6$, 
T.~Nakano$^{11}$, 
N.~Nishi$^7$, 
J.~Nix$^{9}$, 
M.~Nomachi$^{7}$, 
T.~Nomura$^{12}$, 
H.~Okuno$^6$, 
K.~Omata$^6$, 
G.N.~Perdue$^{9}$, 
S.~Perov$^3$, 
S.~Podolsky$^{3, c}$, 
S.~Porokhovoy$^3$, 
K.~Sakashita$^7$, 
N.~Sasao$^{12}$, 
H.~Sato$^8$, 
T.~Sato$^6$, 
M.~Sekimoto$^6$, 
T.~Shinkawa$^{10}$, 
Y.~Sugaya$^7$, 
A.~Sugiyama$^2$, 
T.~Sumida$^{12}$, 
Y.~Tajima$^8$, 
Z.~Tsamalaidze$^3$, 
T.~Tsukamoto$^{2, *}$, 
Y.~Wah$^9$, 
H.~Watanabe$^{9, a}$, 
M.~Yamaga$^{6, d}$, 
T.~Yamanaka$^7$, 
H.Y.~Yoshida$^8$, and 
Y.~Yoshimura$^6$
\\}

\affiliation{
\PUSAN \\
\SAGA \\
\DUBNA \\
\SOKENDAI \\
\TAIWAN \\
\KEK \\
\OSAKA \\
\YAMAGATA \\
\CHICAGO \\
\NDA \\
\RCNP \\
\KYOTO \\
}

\date{\today}

\begin{abstract}
The first dedicated experiment for the rare kaon decay $\kzpinu$ has been performed 
by the E391a collaboration at the KEK 12-GeV proton synchrotron. 
A new upper limit of $\brfinal$ at the 90~\% confidence level was set 
for the branching ratio of the decay $\kzpinu$ using about 10~\% of 
the data collected during the first period of data taking.
\end{abstract}


\maketitle

The rare decay $\kzpinu$ is a Flavor Changing Neutral Current (FCNC) process from 
strange to down quarks and is caused by direct CP violation~\cite{Laur, buras}. 
The theoretical uncertainty in the branching ratio is only 1-2\%, 
while the branching ratio is predicted to be 
$(2.8 \pm 0.4) \times 10^{-11}$ \cite{BGHN06}  
based on the currently known parameters from other experiments. 
The decay is considered an ideal process to study the quark flavor physics and 
a critical test of the standard model as well as a search for new physics beyond it~\cite{isidori}. 
The present experimental limit is 5.9$\times$10$^{-7}$ 
at the 90~\% confidence level~\cite{ktev}; 
the Dalitz decay mode $\pizero \to e^+e^-\gamma$ 
for the final state of $\kzpinu$ was used in the search. 

The E391a experiment at the KEK 12-GeV proton synchrotron was proposed 
to be the first dedicated experiment for the $\kzpinu$ decay and 
aimed to improve the experimental sensitivity by orders of magnitude,  
and to verify the experimental method for 
the next higher sensitivity experiment~\cite{proposal}.
The E391a experiment had three data-taking runs in 2004 and 2005. 
The first data taking (Run-1) was performed from February to June 2004~\cite{thesis}.  
In this Letter, we report results obtained from an analysis of 
about 10 $\%$ of the data collected in Run-1.

The signature of the $\kzpinu$ decay is 2$\gamma$ + {\it nothing} in the final state. 
The energies and hit positions of two photons were measured with an electromagnetic calorimeter. 
The $\pizero$ vertex, $\zvtx$, and its transverse momentum with respect to the beam axis, $\pt$, 
were measured assuming that two photons were produced in 
a $\pizero \to \gamma\gamma$ decay on the beam axis. 
One of the crucial tools for this detection method is 
a neutral beam with small diameter (called ``pencil beam'') 
in order to minimize uncertainty in the $\zvtx$ and $\pt$ measurements. 
Good beam collimation with little halo is also important 
to minimize $\pizero$ production via interactions with detector material. 
The second tool is a hermetic veto system covering the decay region 
to reject background decay modes with additional charged particles and photons.  
High efficiency can be obtained by 
lowering the energy thresholds~\cite{photonveto, chargedveto}. 
The third tool is a decay region with a high vacuum 
to reduce $\pizero$ produced by neutrons in the beam 
interacting with the residual gas. 

\begin{figure*}[t]
  \includegraphics*[width=0.7\textwidth,clip]{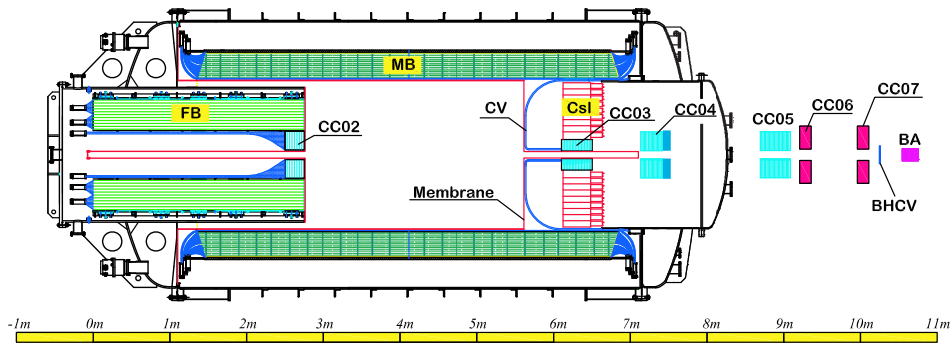}
 \caption{
  Cross section of the E391a detector. $\klong$'s enter from the left side.
 }
\label{fig:det_all}
\end{figure*}

Neutral kaons were produced by striking protons from the 12 GeV proton synchrotron 
on a 60 mm long platinum target.   
The neutral beam was collimated 
to a circular shape having 2 mrad of a half cone angle 
at an angle of 4$^{\circ}$ with respect to the primary proton line~\cite{beamline}.

Figure~\ref{fig:det_all} shows the E391a detector. 
The ``0 m'' in the figure 
corresponds to 11 m downstream from the target. 
Detectors were cylindrically assembled along the beam axis, 
and most of them were installed in a large vacuum vessel 
to eliminate dead material in front of the detectors. 
The fiducial region for $\klong$ decays was from 300 cm to 500 cm. 

The electromagnetic calorimeter was placed at 
the downstream end of the decay region, 
and was 1.9 m in diameter with a 12  $\times$ 12 cm$^2$ beam hole at the center. 
It consisted of 576 blocks of undoped CsI crystal, 
with a size of 7 $\times$ 7 $\times$ 30 cm$^3$ (=16$X_0$) 
except for the inner 24 crystals that were 5 $\times$ 5 $\times$ 50 cm$^3$ (=27$X_0$)~\cite{csi}. 
The energy resolution was $\sigma_E/E \simeq 1\% / \sqrt{E} \oplus 1\%$, 
where $E$ is in GeV, as measured with 25 CsI blocks and a positron beam. 
The average position resolution was 5 mm for photons.
A group of plastic scintillation counters (CV) were placed 
in front of the CsI calorimeter to veto decays involving charged particles. 
The decay region was surrounded by two large lead-scintillator 
sandwich counters, MB and FB. Total thickness of MB (FB) was 13.5 $X_0$ (17.2 $X_0$). 
Their signal was read through wave-length-shifting fibers 
by phototubes (PMTs) with high quantum efficiency~\cite{pmt}. 
The light yield of MB and FB was monitored with 
an LED calibration system. 
The gain shift between the on-beam period and the off-beam period 
was less than 1~\% for both MB and FB. 
Multiple collar-shaped veto counters, CC02 - CC07, were placed along the beam axis 
to detect photons going through the beam hole. 
CC02 was a shashlik type lead-scintillator sandwich counter and  
CC03 was a tungsten-scintillator counter with the layers set parallel to the beam. 
CC04 and CC05 were lead-scintillator counters 
with the layers set perpendicular to the beam. 
CC06 and CC07 consisted of SF-5N lead-glass blocks. 
Back Anti (BA) was located at the end of the beam 
in order to detect photons going through the beam hole without being detected by other detectors. 
BA consisted of lead-scintillator layers and quartz layers, 
and had 14 $X_0$ in total. 
In front of BA, a set of 1 mm thick plastic-scintillators, BHCV, was placed  
to detect charged particles going through the beam hole. 

The vacuum tank was divided internally into two regions by a 20 mg/cm$^2$ 
thick sheet called ``membrane'' to protect the high vacuum region from 
out-gassing from detector components. 
The pressure of the region with detector components was less than 1 Pa, 
and the pressure in the decay region was $1 \times 10^{-5}$ Pa. 

For the CsI calorimeter and all the photon veto detectors, 
we continually calibrated the energy scale factor using cosmic ray muons and 
minimum ionization particles in the beam during the operation. 
We also studied the energy scale factor of the CsI calorimeter 
using the special data, in which 
an aluminum(Al) target with a thickness of 5 mm was inserted 
in the beam at $z = 280.5$ cm (downstream of CC02) 
in order to produce $\pi^0$ by neutrons in the beam. 
We adjusted the gain of each CsI crystal so that the invariant mass 
of two photons agree with the $\pi^0$ mass. 
This data was also used to check our 
event reconstruction performance. 
The more detailed description of the calibration 
can be found in~\cite{thesis, csi}. 

%
%
The trigger was designed to accept $\kzpinu$, $\kzpipi$ and $\kzpipipi$ decays. 
The latter two decays were used 
to study the detector response, and the $\kzpipi$ decays were 
used to measure the number of $\klong$ decays. 
For triggering purpose, 
we grouped eight neighboring CsI crystals and defined 72 regions in total. 
The trigger required that there were two or more such regions 
with $\ge$ 60 MeV energy deposit in each. 
The trigger also required 
no energy deposit in CV and several photon veto detectors. 
For example, the veto threshold for the total energy deposit in MB was 15 MeV. 
The trigger rate was 800 events per 2 second beam delivery with 
a typical intensity of $2.5 \times 10^{12}$ protons on the target. 
The live time ratio was 78~\% with 
a network distributed data acquisition system with multiple CPUs. 
The electronics and data acquisition system is briefly 
described elsewhere~\cite{thesis, csi}. 

In the offline analysis, 
we first looked for photons in the CsI calorimeter. 
Each cluster of energy deposits was required to have 
the transverse shower shape consistent with a single electromagnetic shower. 
The effective energy threshold of each cluster was 10 MeV. 
We assumed that the two photons came from a $\pizero$ decay, and 
reconstructed $\zvtx$  
requiring the two photon invariant mass to have the $\pizero$ mass. 

We selected events with exactly two photons hitting the CsI calorimeter 
and applied selection criteria (cuts) to suppress background events. 
In Run-1, the downstream membrane was partially hanging in the beam by error, 
at $z=550$ cm. 
This produced 
a large number of background events 
because neutrons in the beam core struck the membrane and produced 
secondary $\pizero$'s. 
If multiple $\pizero$'s were produced at the membrane, 
and two photons from different $\pizero$'s were detected 
(``core neutron multi-$\pizero$'' event), 
it became a serious background event because we were not able to reconstruct $\zvtx$ correctly  
and these events were distributed 
in the fiducial region. 
On the other hand, these events had extra photons 
in the final state, and thus can be suppressed by detecting those extra photons. 

\begin{figure}
 \includegraphics*[width=0.5\textwidth,clip]{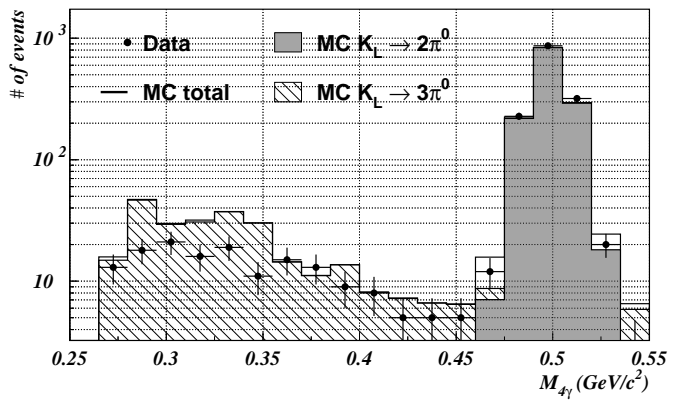}
 \caption{%
 Distribution of the invariant mass of four photons, $M_{4\gamma}$, with 
the cuts on the photon veto detectors and the shower shape of photons 
in the CsI calorimeter. 
The dots show data, the open solid histogram shows total MC, 
the closed solid histogram shows  $\kzpipi$ MC and 
the hatched histogram shows  $\kzpipipi$ MC. 
 }%
 \label{fig:g4mass_plot}
\end{figure}

In order to suppress events involving extra photons, 
we required energy deposit in each photon veto detector to be less than 
the threshold listed in Table~\ref{tbl:threshold}. 
The rejection power of the photon veto was evaluated with 
four-photon event samples from $\kzpipi \to 4\gamma$ and 
$\kzpipipi \to 6\gamma$ with two missing photons. 
Figure~\ref{fig:g4mass_plot} shows the invariant mass 
of four photons, $M_{4\gamma}$, 
after applying all the cuts on the photon veto detectors. 
With all the photon veto cuts, 
the ratio of the number of $\kzpipi$ events in 
$0.45 \le M_{4\gamma} \textrm{(GeV/c$^2$)} \le 0.55$ 
to the number of $\kzpipipi$ events in $M_{4\gamma} \textrm{(GeV/c$^2$)} \le 0.45$ 
improved by a factor of 11.
This improvement was consistent with 
the expectation of 
GEANT-3 based~\cite{ref:geant3} Monte Carlo simulation (MC) within 18~\%.

\begin{table}[htb]
\begin{center}
\begin{minipage}{0.5\textwidth}
\caption{%
List of the thresholds applied to the photon veto detectors. 
$E_Q$ is the total light yield in the BA quartz layers, and 
$E_S$ is the total energy deposit in the BA scintillator layers. 
The signal efficiency for a cut $A$, $\sigeff$, 
is the ratio of the number of events with all cuts 
to the number of events with all cuts except for the cut $A$.  
We estimated $\sigeff$ using MC $\kzpinu$ events except for the cut on BA. 
For BA, we first evaluated $\sigeff$ for $\kzpipi$ and $\kzpipipi$ decays using real data, 
which were $0.638 \pm 0.022\stat$ and $0.658 \pm 0.022\stat$, respectively, 
and then assigned the average as $\sigeff$ for $\kzpinu$ signal.
}%
\label{tbl:threshold}
\begin{tabular}{cccccc}
\hline\hline
Detector & threshold & \phantom{x} $\sigeff$ \phantom{x} & 
Detector & threshold & \phantom{x} $\sigeff$ \phantom{x}\\
\hline 
CC02 & 4 MeV & 1.0 & CC06 & 5 MeV & 0.98\\
\hline
CC03 & 1.5 MeV & 0.98 & CC07 & 50 MeV & 0.99\\
\hline
CC04 & 3 MeV &  0.98 & FB & 2 MeV & 0.91\\
\hline\hline
\end{tabular}\\

\vspace*{2ex}

\begin{tabular}{ccc}
\hline\hline
Detector & threshold & \phantom{x} $\sigeff$ \phantom{x} \\
\hline
CsI  & 3 MeV for the CsI crystals which do& \\
       & not belong to the photon clusters& 0.78\\
\hline
MB   & 1 MeV for the inner modules, and  & \\
     & 0.5 MeV for the outer modules & 0.60\\
\hline
BA   & 0.5 MIP for $E_{Q}$, and & \\
     & $E_{Q}/E_{S} \ge 10$  & 0.65\\
\hline\hline
\end{tabular}
\end{minipage}
\end{center}
\end{table}

From the MC study, we found that 
the ``core neutron multi-$\pizero$'' events had low $\pt$ and 
were populated at the downstream $\zvtx$ region. 
In order to minimize the number of such background events, 
we used this characteristic and required a parameter,  
$\alpha \; \equiv \; P_T \textrm{(GeV/c)}- 8.0 \times 10^{-4} \textrm{(GeV/c$\cdot$cm)} 
\times \zvtx \textrm{(cm)}\quad (\zvtx < 525 \textrm{cm})$, 
to be larger than $-0.225$ GeV/c. 
Another cut, $\pt \ge 0.12$ GeV/c, was applied to suppress $\kzgg$ background 
and $\Lambda \rightarrow \pi^0 \;n$ background, whose maximum $P_T$ is 0.109 GeV/c. 
The upper boundary on $\pt$ was determined to be $\pt \le 0.24$ GeV/c 
from the kinematical limit of the $\kzpinu$ decay ($P_{max} = 0.231$ GeV/c), 
allowing for the smearing effect due to detector resolutions.

\begin{figure}
 \includegraphics*[width=0.5\textwidth,clip]{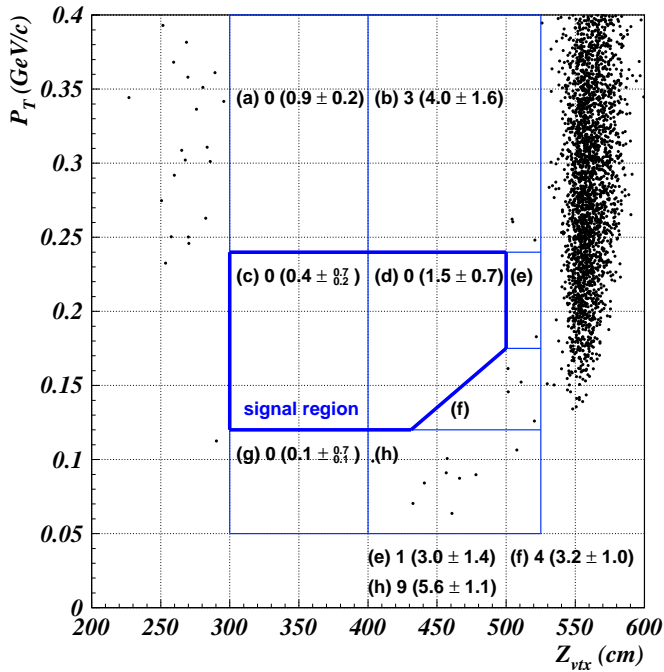}
 \caption{%
 $\zvtx$ versus $\pt$ with all the event selection cuts. 
 The number of observed (total expected background) events are shown.
 The expected number of background events was consistent with 
 the observed number of events for all the regions. 
 }%
 \label{fig:final_plot}
\end{figure}

After applying all the selection cuts, 
we estimated the number of remaining background events in the eight 
$\pt$--$\zvtx$ regions with the signal regions (c) and (d) 
as shown in Fig.~\ref{fig:final_plot}. 
Except for the regions (a), (c) and (g), 
the dominant background source was the ``core neutron multi-$\pizero$'' event. 
We evaluated the number of ``core neutron multi-$\pizero$'' events 
using a relational expression with two independent selection cuts:  
$N_{bkg} = N' \times \textrm{(cut-1 rejection)} \times \textrm{(cut-2 rejection)}$, 
where ``cut-1'' is a set of cuts on CV, MB, CC03, CC04, CC06 and CC07, 
``cut-2''  is a set of cuts on the cluster energy and the cluster hit position,   
and $N'$ is the number of events with all the selection cuts except for the cut-1 and the cut-2. 
We checked that (i) $(97 \pm 3)$~\% of the $N'$ was the ``core neutron multi-$\pizero$'' 
events even without the cut-1 and cut-2, 
and (ii) the cut-1 and the cut-2 were independent of each other\footnote{%
For each selection cut in the cut-2, we examined the ratio of the number of 
events passing the cut to the number of events failing the cut. 
The ratio for the cluster energy cut was 
$0.79 \pm 0.12$ with the cut-1 and  
$0.73 \pm 0.03$ without the cut-1.  
The ratio for the cluster hit position cut was 
$(5.1 \pm 2.6) \times 10^{-2}$ with the cut-1 and  
$(5.2 \pm 0.6) \times 10^{-2}$ without the cut-1.
}. 
The number of ``core neutron multi-$\pizero$'' background events 
was $0.0^{+0.7}_{-0.0}$ in region (c) and $1.5 \pm 0.7$ in region (d). 
The number of background events caused by the halo neutrons interacting with the detector 
material (CC02) and producing one or more $\pizero$ was $0.9 \pm 0.2 $ in region (a) 
and  $0.04 \pm 0.04$ in region (c). 
The background events caused by the core neutrons interacting with 
the membrane and producing $\eta$'s 
(``core neutron $\eta$'' events) was reconstructed 
around region (c) because the $\pizero$ mass was assumed. 
For all the events, we re-calculated the decay vertex assuming $\eta$ mass ($Z_{\eta}$) 
and then rejected events around the membrane in the beam, $525 \le Z_{\eta} \textrm{(cm)} \le 575$. 
The remaining number of ``core neutron $\eta$'' events was $0.4 \pm 0.2$, 
which was the largest component in region (c).  
The background events from $\kzpipi$ with two missing photons 
was evaluated with MC. 
The number of $\kzpipi$ background events in the signal region 
was $0.04 \pm 0.03$,  
where the error includes the MC statistics and the systematic uncertainties, 
of which the dominant source was 
the mismatch between data and MC in the transverse shower shape of photon in the CsI calorimeter. 
Moreover, the $\kzpipi$ background events were the largest component in region (g). 
The total number of background events in the signal region was 
estimated to be $0.4^{+0.7}_{-0.2}$ in region (c) and 
$1.5 \pm 0.7$ in region (d).

\begin{figure}
 \includegraphics*[width=0.5\textwidth,clip]{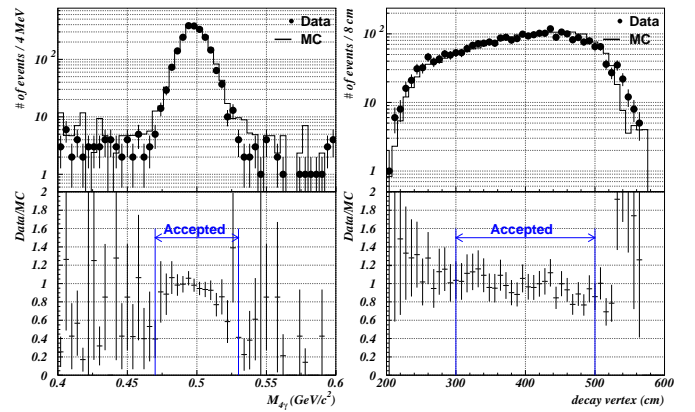}
 \caption{%
Distribution of the invariant mass (left) and the decay vertex (right) 
for the $\kzpipi$ decays. In the top plot, the dots show the data 
and the histogram shows the MC. 
The bottom plot shows the ratio of the data to the MC.
 }%
 \label{fig:norm-plot}
\end{figure}

We estimated the acceptance of $\kzpinu$ decay 
to be $(0.657 \pm 0.016) \times 10^{-2}$ 
based on cut efficiencies evaluated with the real data and MC study. 
The main components of the acceptance loss were
the cuts on MB and BA photon veto detectors.  
In order to estimate the number of $\klong$ decays in this search, 
we analyzed $\kzpipi$ decays. 
The invariant mass and the reconstructed decay vertex 
for  $\kzpipi$ are shown in Fig.~\ref{fig:norm-plot}. 
In the $\kzpipi$ signal region: 
$0.47 \le M_{4\gamma} \textrm{(GeV/c$^2$)}\le 0.53$,  
and $300 \le \zvtx \textrm{(cm)} \le 500$, 
there were 2081 $\kzpipi$ events after subtracting 
30 $\kzpipipi$ background events. 
Based on the MC study, we estimated that the acceptance of $\kzpipi$ decay 
was $1.41 \times 10^{-3}$. 

The different final states between the signal and normalization modes 
caused systematic uncertainties in the single event sensitivity. 
We assigned the total systematic uncertainty in the single event sensitivity to be $\pm$~7.0~\%. 
The large sources of systematic uncertainty came from the mismatch 
between data and MC in the transverse shower shape of the photon (4~\%)
and the energy distribution in MB (4.2~\%).

With the $\kzpipi$ branching ratio, $(8.83 \pm 0.08) \times 10^{-4}$~\cite{PDG}, 
we estimated the number of $\klong$ decays to be $\nklong$. 
The single event sensitivity was $\sesfinal$. 
Since we observed no events in the signal region, 
we set a new upper limit on the branching ratio of $\kzpinu$ to be 
$< \brfinal$ at the 90~\% confidence level based on the Poisson statistics. 
This represents an improvement of a factor of 2.8 over the current limit~\cite{ktev}.

We are grateful to the operating crew of the KEK 12-GeV proton synchrotron 
for their successful beam operation during the experiment. 
We express our sincere thanks to Professors H.~Sugawara, Y.~Totsuka, M.~Kobayashi 
and K.~Nakamura for their continuous encouragement and support. 
Thanks are due to A.J.~Buras and G.~Isidori for useful theoretical discussions and encouragement. 
This work has been partly supported by a Grant-in-Aid from the MEXT and JSPS in Japan, 
a grant from NSC in Taiwan and the U.S. Department of Energy.

\vspace*{1ex}
\noindent  
$^*$Deceased \\
$^a$Present address: KEK, Tsukuba, Ibaraki, 305-0801 Japan. \\
$^b$Also Institute for High Energy Physics, Protvino, Moscow region, 142281 Russia. \\
$^c$Also Scarina Gomel' State University, Gomel', BY-246699, Belarus. \\
$^d$Present address:  Osaka University, Toyonaka, Osaka, 560-0043 Japan.

\bibliographystyle{plain}

\end{document}